\documentclass[showpacs,preprintnumbers,amsmath,amssymb,prb,preprint,superscriptaddress]{revtex4}

\usepackage{graphicx}
\usepackage{dcolumn}
\usepackage{bm}

\begin{document}

\title{Thermally induced rotons in two-dimensional dilute 
Bose gases}

\author{Flavio S. Nogueira}
\email{nogueira@physik.fu-berlin.de}
\affiliation{Institut f{\"u}r Theoretische Physik,
Freie Universit{\"a}t Berlin, Arnimallee 14, D-14195 Berlin, Germany}
\author{Hagen Kleinert}
\email{kleinert@physik.fu-berlin.de}
\affiliation{Institut f{\"u}r Theoretische Physik,
Freie Universit{\"a}t Berlin, Arnimallee 14, D-14195 Berlin, Germany}

\date{Received \today}

\begin{abstract}
We show that roton-like
excitations are thermally induced in a
two-dimensional dilute Bose gas
as a
consequence of the strong phase fluctuations in two dimensions.
At low momentum, the roton-like excitations lead for small enough 
temperatures to an anomalous 
phonon spectrum with a temperature dependent exponent reminiscent of 
the Kosterlitz-Thouless transition. Despite the anomalous form of the 
energy spectrum, it is shown that the corresponding effective theory of 
vortices describes the usual Kosterlitz-Thouless transition. The possible 
existence of an anomalous normal state in a small temperature interval is 
also discussed.
\end{abstract}


\maketitle

\section{Introduction}

Superfluidity in interacting Bose systems has been a fascinating
research topic for several decades. After the achievement of
Bose-Einstein condensation in dilute atomic gases,\cite{BEC} many
of the remarkable properties of superfluid $^4$He were also
observed in these weakly interacting systems, such as
vortices and
Josephson oscillations.\cite{Pita} In order
to understand their
superfluid properties,
it is necessary to clarify the role
 of the
elementary excitations in determining these properties.
According to a celebrated criterion due to Landau, a
 Bose
gas without interactions cannot be a superfluid, since its spectrum
$\varepsilon_p=p^2/2m$ makes it indifferent to Galilei
transformations. A superfluid, on the other hand,
is resistant to a slowdown of the molecules
due to the loss of Galilei invariance.
A Bose system with the relativistic looking excitation  spectrum
 $E_p=c p$ has this property, thus being a superfluid
by Landau's argument. Thus,
phonon-like excitation spectra are
an
essential part of a
superfluid.

In superfluid helium, which is a strongly interacting
Bose system, the interatomic potential
has an attractive short-range part
over a distance $a_0$,
the average interatomic
distance.
This sets the scale of
a further important set of excitations.
Scattering experiments of  neutrons  show that around
a momentum $p_0\approx1/a_0$
the spectrum behaves like $E_p=\Delta+(p-p_0)^2/2m^*$.
These elementary excitations are called {\rm rotons}.

Feynman \cite{Feyn} was the first to recognize
the importance of the
rotons for a superfluid.
In modern language, Feynman's theory
describes rotons as a result
of
large
{\it quantum}
 phase
fluctuations at low temperatures. These produce
small vortex loops of size $\sim a_0$. At higher temperatures
where thermal fluctuations take over,
the vortex loops combine to larger {\it thermally} excited vortex
loops which profit from the high
configurational entropy
of line-like excitations.
At the critical $ \lambda $-point,
these  loops become infinitely long and destroy the order
of the superfluid.\cite{Kleinert}

In two dimensions, phase fluctuations are so strong that
they destroy the
long-range order at any temperature.\cite{HMW}  As noted first by
Kosterlitz and Thouless (KT), 
\cite{KT} there still exists
a phase transition driven by phase fluctuations.
At low temperature, a film of superfluid
$^4$He contains vortex- antivortex pairs bound by Coulomb attraction,
whose
 unbinding causes the KT transition.
One of the most important predictions of this theory is
the universal jump to zero of the 
superfluid stiffness at the
critical temperature.\cite{NelKost} This
destroys
the
excitations of energy $E=cp$, and thus the superfluidity.

The increased relevance of phase fluctuations in two
 dimensions suggests that Feynman's
rotons are more abundant
than in three dimensions.
We want to argue that
this is indeed the case:
in spite of the weak interactions,
a dilute Bose gas
possesses roton-like excitations
which are the precursors of the high-temperature vortices that 
lead to a KT phase transition.
This is quite
remarkable since, contrary to superfluid
helium, the bare interaction does not contribute with an attractive part 
to fix the size of $p_0$.

The plan of the paper is as follows. In Sect. II we briefly review the known 
results of the $t$-matrix formalism in $d$ dimensions 
for dilute Bose gases. Sect. III contains most of the main results of 
the paper. There we apply the so called dielectric formalism \cite{RPA,Kondor} 
to the two-dimensional dilute Bose gas. We will essentially work out a 
RPA approximation in two dimensions. Since the 
dielectric formalism is well known in three-dimensional 
applications, we will not discuss its derivation here, referring the reader 
instead to the literature. The focus will be in the application of 
the method to two dimensions. From our calculations a new, temperature dependent, 
excitation spectrum emerges, namely, 
\begin{equation}
E_p=\sqrt{\varepsilon_p^2+2{g_2}n\varepsilon_p\left[1-\frac{Tm}{\pi n}\ln(pa)\right]}.
\end{equation}
We will show that the above spectrum allows for thermally excited roton-like 
excitations and that its low momentum behavior exhibits an anomalous 
power behavior with a temperature dependent exponent $\eta_0(T)$. 
Remarkably, in the two-dimensional dilute Bose gas at finite temperature, 
the RPA analysis will lead to closed form analytic results including roton-like 
excitations.  
In Sect. IV we 
discuss the effect of phase fluctuations in the system. Our analysis in this 
Section will allow us to derive the approximate   
critical temperature of the system as $\eta_0(T_c)=1/4$. The actual critical 
temperature is determined in the usual way 
following the Kosterlitz-Thouless vortex unbinding mechanism.\cite{KT} 
It will also be shown that a crossover 
temperature $T_*$ exists, above which our anomalous spectrum becomes unstable. 
Sect. V concludes the paper. The whole discussion is supplemented with 
two Appendices containing calculational details.  

\section{The effective interaction at zero temperature}

In general  
the effective interaction for any dimension $d<4$ depends 
on a momentum scale $\bar{p}$ associated to the energy of the scattered 
particles. At zero temperature the effective interaction 
in $d$ dimensions reads
\begin{equation}
\label{coupling1}
g(\bar{p})=\frac{4\pi^{d/2}a^{d-2}/m}{2^{2-d}\Gamma(1-d/2)(\bar{p}a)^{d-2}
+\Gamma(d/2-1)},
\end{equation}
where $m$ is the reduced mass of the scattered particles and 
$a$ is the $s$-wave scattering length. The 
above result is essentially exact at zero temperature and corresponds  
to a geometric sum of ladder diagrams 
which gives the only nonvanishing contribution to the vertex function.\cite{Uzunov} 

From Eq. (\ref{coupling1}) we derive the 
exact renormalization group (RG) $\beta$-function for the 
dimensionless coupling $\tilde{g}(\bar{p})=m\bar{p}^{d-2}g(\bar{p})$
\begin{equation}
\label{betag}
\beta(\tilde{g})\equiv\bar{p}\frac{\partial\tilde{g}}{\partial\bar{p}}
=(d-2)\left[\tilde{g}+\frac{d~\Gamma(-d/2)}{2^{d+1}\pi^{d/2}}
\tilde{g}^2\right].
\end{equation}
For $2\leq d<4$ the only fixed point is $\tilde{g}_*=0$. 
On the other hand, for $d<2$ a nontrivial fixed point 
located at 
\begin{equation}
\label{fixpoint}
\tilde g_*=-\frac{2^{d+1}\pi^{d/2}}{d\Gamma(-d/2)}
\end{equation}
exists. Thus, $d=2$ is the upper critical dimension for the 
dilute Bose gas. 
Writing $d=2-\epsilon$ and expanding Eq. (\ref{fixpoint}) for 
small $\epsilon$, we obtain
\begin{equation}
\tilde g_*\approx 2\pi\epsilon.
\end{equation}
In the dimension interval $2<d<4$ we can easily take the limit 
$\bar p a\to 0$ in Eq. (\ref{coupling1}) to obtain
\begin{equation}
\label{coupling}
g_0=\frac{4\pi^{d/2}a^{d-2}}{\Gamma(d/2-1)m}.
\end{equation}
For $d=3$ the above equation reproduces the familiar formula  
$g_0=4\pi a/m$. For $d=2$, however, we cannot set $\bar{p}a=0$.  
In two dimensions Eq. (\ref{coupling1}) becomes

\begin{equation}
\label{coupling2D}
g_{\rm 2D}(\bar{p})=\frac{2\pi/m}{\ln 2-\gamma-\ln(\bar{p}a)},
\end{equation}
where $\gamma$ is the Euler constant. Eq. (\ref{coupling2D}) 
is in agreement with previous work.\cite{Stoof1,Lee} 

 The first theories for the two-dimensional dilute Bose gas were developed by
Popov \cite{Popov} and Schick.\cite{Schick} More recently, an improved
version of Popov's theory was presented by Stoof and collaborators.\cite{Stoof0,Stoof}
In Popov's approach,\cite{Popov} the bare interaction
$g_0 \delta ({\bf x})$  is replaced by an effective
interaction $g\delta ({\bf x})$ determined by a $t$-matrix,
leading in $d$ dimensions
with $d\in [2,4)$ to \cite{FH}
\begin{equation}
\label{coupling2}
g_d=\frac{4\pi^{d/2}a^{d-2}/m}{2^{2-d}\Gamma(1-d/2)(na^d)^{d-2}
+\Gamma(d/2-1)}.
\end{equation}
where
$n$ is the density and
$a$ the $s$-wave scattering
length. The above interaction corresponds precisely to the one 
given in Eq. (\ref{coupling1}), where we have set $\bar p a=na^d$. 
For $d\rightarrow 2$ we obtain the two-dimensional coupling 
constant of the Popov-Schick theory \cite{Popov,Schick}
\begin{equation}
\label{coupling3}
{g_2}\equiv \lim_{d\to 2}g=-\frac{2\pi/m}{\ln(e^\gamma na^2/2)},
\end{equation}
where $\gamma$ is the Euler-Mascheroni constant.
The logarithm in the denominator implies that the
effective repulsion decreases only very slowly
with decreasing density.\cite{FH}
Fisher and Hohenberg \cite{FH}
have shown within
the Popov-Schick theory that
the dilute limit $na^d\ll 1$ of the $d>2$ theory must
be replaced for $d=2$  by $\ln\ln(1/na^2)\gg1$.

\section{The dielectric formalism in $d=2$}

The $t$-matrix result (\ref{coupling1})
incorporates, via a Lippmann-Schwinger integral equation,
 the sum of all ladder diagrams.
In this Section we take into account the sum of bubble diagrams
of the plasmon type,\cite{FPH} which are nonvanishing 
at finite temperature.
This corresponds to the
random phase approximation (RPA), which sums
geometrically the
diagrams shown in Fig.~1. This approximation has often  been
applied in $d=3$ dimensions.\cite{RPA,Kondor}
In RPA the vertex function containing the effects of
both resummations is given 
explicitly in $d$ dimensions by
\begin{equation}
\label{vertex}
\Gamma(i\omega,{\bf p})=\frac{g_d}{1-g_d\widetilde \Pi(i\omega,{\bf p})},
\end{equation}
where the polarization bubble is given by
\begin{equation}
\label{Pi}
\widetilde \Pi(i\omega,{\bf p})=\!
\frac{1}{\beta}\!\sum_{n}\!
\int\!\frac{d^dq}{(2\pi)^d}G_0(i\omega+i\omega_n,
{\bf p}+{\bf q})G_0(i\omega_n,{\bf q}),
\end{equation}
with $G_0(i\omega,{\bf p})=1/(i\omega-\varepsilon_p)$. 
The chemical potential is canceled by the Hartree contribution.\cite{FH}
\begin{figure}
\includegraphics[width=12cm]{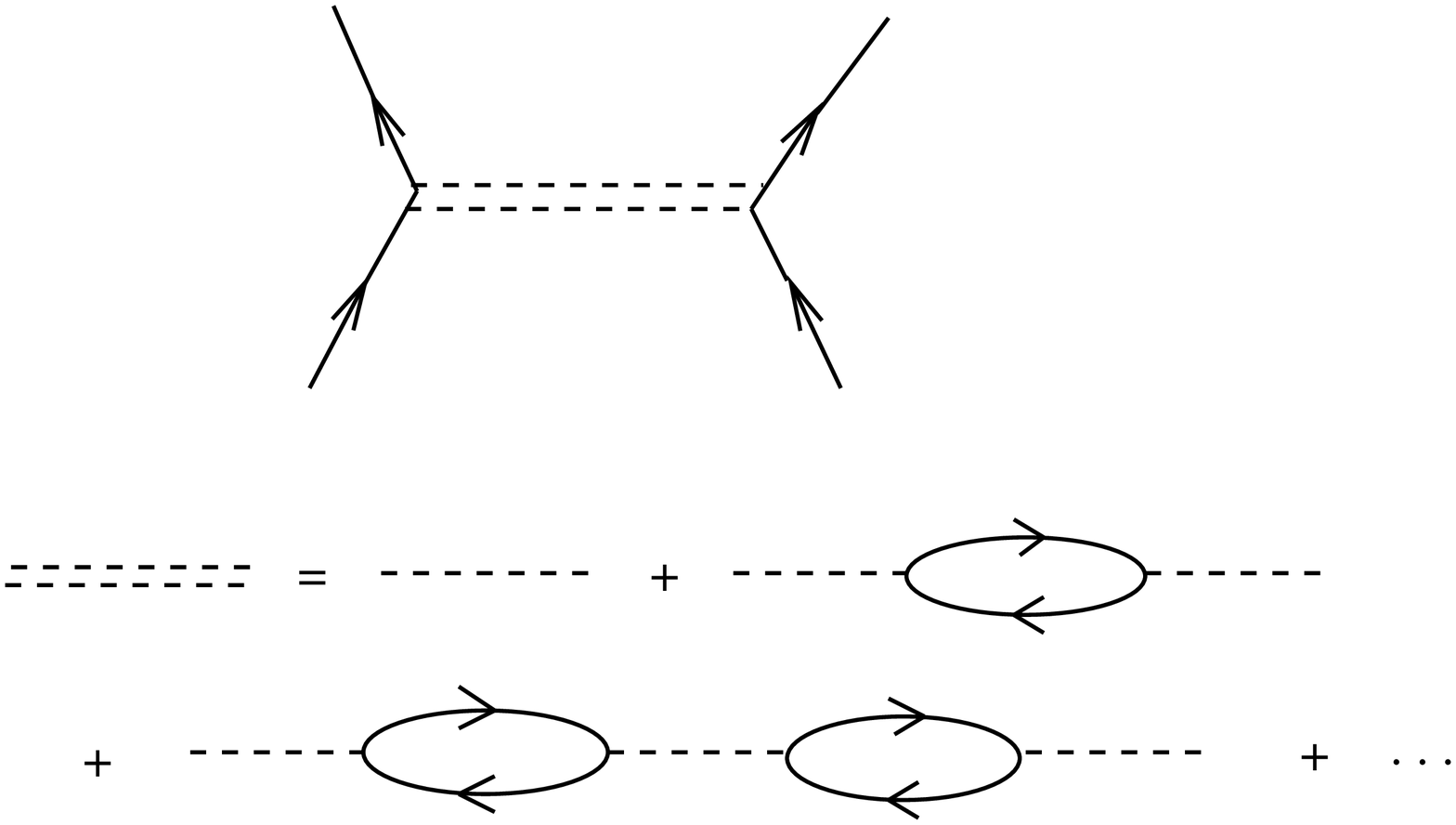}
\caption{Feynman diagram representation of the vertex function
Eq. (\ref{vertex}). The dashed line represents the bare
interaction. The vertex function is obtained as a geometric
series of polarization bubbles.}
\end{figure}
The denominator of (\ref{Pi}) determines the {\it regular} part 
$\epsilon_r(i\omega,{\bf p})$ of the dielectric function 
$\epsilon(i\omega,{\bf p})$ renormalizing  
the density correlation function, i.e.,
\begin{equation}
\chi_{\rho\rho}(i\omega,{\bf p})=\frac{\Pi(i\omega,{\bf p})}
{\epsilon(i\omega,{\bf p})}
\end{equation}
where
\begin{equation}
\label{dielconst}
\epsilon(i\omega,{\bf p})=1-g_d[\Pi_0(i\omega,{\bf p})+\widetilde{\Pi}
(i\omega,{\bf p})],
\end{equation}
with
\begin{equation}
\label{Pi0}
\Pi_0(i\omega,{\bf p})=\frac{2n\varepsilon_p}{(i\omega)^2-\varepsilon_p^2},
\end{equation}
contributing to the singular part of the dielectric function.\cite{Kondor} 
The regular contribution, being given by a particle-hole diagram,  
vanishes at zero temperature. 
We will treat $\widetilde{\Pi}(i\omega,{\bf p})$ in 
the so called classical approximation, where
the Bose-Einstein distribution function
$n_B(x)\equiv 1/(e^{\beta x}-1)$ is replaced by
 $n_B(x)\approx T/x$. Thus, the classical 
approximation is valid in the limit $p\lambda_T\ll 1$, 
where $\lambda_T=(2mT)^{-1/2}$ is the thermal wavelength. Therefore, 
we obtain in the limit  $p\lambda_T\ll 1$ the result 
\begin{equation}
\label{Pi1}
\widetilde{\Pi}(i\omega,{\bf p})=-A_dm^2Tp^{d-4}
\left[e^{i\pi(d-2)/2}\left(1-{i\omega}/{\varepsilon_p}\right)^{d-3}+
e^{-i\pi(d-2)/2}\left(1+{i\omega}/{\varepsilon_p}\right)^{d-3}\right],
\end{equation}
with $A_d=2^{2-d}\pi^{-d/2}\Gamma(d/2-1)\Gamma(3-d)$.
A derivation of Eq. (\ref{Pi1}) using Feynman parameters is given 
in the Appendix.
For $d=3$, this reduces to the well-known result
\cite{RPA,Kondor}
\begin{equation}
\label{Pi1-3d}
\widetilde{\Pi}(i\omega,{\bf p})\mathop{=}^{d=3}
-i\frac{Tm^2}{2\pi p}\ln\left(
\frac{i\omega+\varepsilon_p}{i\omega-\varepsilon_p}\right).
\end{equation}

In $d=2$ dimensions Eq. (\ref{Pi1}) has a pole
 associated with a logarithmic short
distance divergence. We will remove this divergence using the $s$-wave 
scattering length as short-distance cutoff. This leads to
\begin{equation}
\label{Pi1-2d}
\widetilde{\Pi}(i\omega,{\bf p})\mathop{=}^{d=2}-\frac{2mT}{\pi}\frac{\varepsilon_p\ln(pa)}{(i\omega)^2-\varepsilon_p^2}.
\end{equation}
The spectrum of elementary excitations is obtained from the
poles of $\chi_{\rho\rho}(i\omega,{\bf p})$.\cite{Kondor,Gavoret,HM}
At zero temperature,
 only $\Pi_0(i\omega,{\bf p})$ contributes,
 and we recover the usual Bogoliubov spectrum. 

The excitation spectrum is obtained from the pole of the density correlation 
function, which corresponds to the vanishing of the dielectric constant 
(\ref{dielconst}). In this way we obtain the excitation spectrum 
announced in the Introduction:
\begin{equation}
\label{spec-2d}
E_p=\sqrt{\varepsilon_p^2+2{g_2}n\varepsilon_p\left[1-\frac{Tm}{\pi n}\ln(pa)\right]}.
\end{equation}
The above energy spectrum possesses a thermally induced roton-like minimum. 
The excitation spectrum is shown in Fig. 2
for a suitable set of parameters.
\begin{figure}
\begin{picture}(145.87,143.05)
\unitlength.7mm
\put(-18,3){\includegraphics[width=7cm]{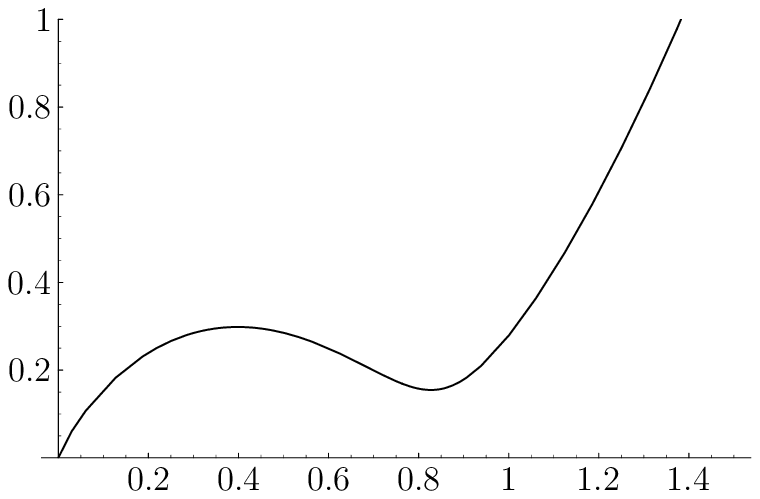}}
\put(80,4.5){$ p $}
\put(-20,60){$ E_p $}
\end{picture}
\caption{The RPA corrected excitation spectrum of Popov-Schick theory
exhibiting a roton-like minimum. The parameters in
a system of units such that $k_B=\hbar=1$ are $m=0.5$ cm$^{-2}$ sec, $n=0.01$ cm$^{-2}$,
$T=1.2$ sec$^{-1}$, and $a=2$ cm.}
\end{figure}

The approximate position of the roton minimum can be determined in the following way using 
the Landau criterion for superfluidity. According to Landau criterion, the 
critical velocity above which excitations appear in the fluid is given 
by the minimal value of the ratio $E_p/p$. This minimum value corresponds 
to the point $p=p_0$ for which
\begin{equation}
\label{vcr-p}
\frac{dE_p}{dp}=\frac{E_p}{p}.
\end{equation}
On the other hand, the energy spectrum (\ref{spec-2d}) satisfies the 
equation:
\begin{equation}
\label{cond-spec}
\frac{dE_p}{dp}=\frac{E_p}{p}+\frac{\varepsilon_p}{pE_p}\left(
\varepsilon_p-\frac{Tmg_2}{\pi}\right).
\end{equation}
The point $p_0$ for which Eqs. (\ref{vcr-p}) and (\ref{cond-spec}) 
coincide is determined by the vanishing of the second term 
on the right-hand side of Eq. (\ref{cond-spec}), i.e., 
\begin{equation}
\label{p0}
p_0(T)=\frac{1}{a_T}=m\sqrt{\frac{2Tg_2}{\pi}}. 
\end{equation}
The above value of $p$ corresponds approximately to the position of the roton-like 
minimum of the excitation spectrum (\ref{spec-2d}). 
The $T$-dependent length scale  $a_T$ nearly replaces
the $a_0$ of bulk $^4$He. 

The consistency of the calculation leading to 
the spectrum (\ref{spec-2d}) can be checked by computing 
the spectrum also from the pole of the anomalous propagator. It is 
well known that the pole of the propagator should be the same 
as the one from the density correlation function,\cite{HM,Kondor} 
although most approximations fail to fulfill this requirement. 
After analytically continuing to real frequencies, we obtain that 
the pole of the propagator is given by the solution of the equation 
\cite{RPA,Kondor}
\begin{equation}
\label{pole-prop}
\omega^2-\varepsilon_p^2-2n\varepsilon_p\Gamma(\omega,{\bf p})=0.
\end{equation}
The above equation is a generalization of the 
Bogoliubov result for the excitation spectrum. Indeed, it 
corresponds to an improvement of the Bogoliubov 
result where the coupling constant 
$g_2$ is replaced by the vertex function (\ref{vertex}).\cite{RPA} 
Solution of Eq. (\ref{pole-prop}) with the polarization 
bubble (\ref{Pi1-2d}) gives precisely the energy spectrum 
(\ref{spec-2d}).    

The excitation spectrum 
(\ref{spec-2d}) has another interesting property in the  
low-momentum regime, which for $mg_2/\pi< 1$ is defined by
\begin{equation}
\label{ineq}
p^2\ll p_0^2(T)=\frac{2m^2Tg_2}{\pi}<\frac{1}{\lambda_T^2}=2mT\ll 4\pi n.
\end{equation}
In the above low-momentum regime 
the logarithm gives rise to
an {\em anomalous\/} power behavior
\begin{equation}
\label{spec-lowp}
E_p\approx \sqrt{\frac{{g}_2n}{m}} a^{-Tm/2\pi n}p^{1-Tm/2\pi n}.
\end{equation}
The exponent of $p$  in (\ref{spec-lowp}) can be rewritten as
\begin{equation}
\label{expon}
\sigma(T)\equiv 1-\frac{Tm^2}{2\pi\rho_s(0)}=1-\eta_0(T),
\end{equation}
where $\rho_s(0)=mn$ is the superfluid mass density at zero temperature,
and $\eta_0(T)\equiv Tm^2/2\pi\rho_s(0)$. Interestingly, precisely the 
same exponent arises in the spin wave treatment of {\it classical} 
phase fluctuations in the study of the KT transition. We will revisit this 
analysis in the next Section,  where it  will also be shown 
that despite the anomalous scaling of the spectrum,  
the KT transition occurs as usual at higher temperatures, 
in agreement with the discussion of Ref.~\onlinecite{Stoof}. This is to be  
expected, since the KT transition actually occurs through a vortex-antivortex 
unbinding mechanism. 

The elementary excitations described by
 Eq. (\ref{spec-lowp}) are stable only for
$\eta_0(T)<1$, i.e., for $T<2\pi\rho_s(0)/m^2\equiv T_*$. Note that 
the low-momentum inequality (\ref{ineq}) already requires 
$T\ll T_*$. But this condition can be softened to just $T<T_*$, in which case 
Eq. (\ref{spec-lowp}) is still approximately valid up to logarithmic 
corrections. The role of the temperature $T_*$ will be discussed later.

It is instructive
to calculate the low-momentum
contribution to the superfluid density following from (\ref{spec-lowp}).
Since vortices are not included in the above calculation, the result
will
show only the anomalous phonon contribution to the
so called background superfluid density, $\rho_{s0}$.
The calculation
is based on
 the  Landau
prescription,\cite{Pita} according to which  the superfluid density is
$\rho_{s0}^{\rm ph}(T)=\rho-\rho_{n0}^{\rm ph}(T)$, where
$\rho_{n0}^{\rm ph}(T)$
is the
anomalous phonon contribution to the normal fluid density. From the 
Landau formula the normal background fluid density is given by
\begin{equation}
\label{rhon}
\rho_{n0}=\frac{\beta}{d}\int\frac{d^dp}{(2\pi)^d}\frac{p^2
e^{\beta E_p}}{( e^{\beta E_p}-1)^2}.
\end{equation}
Now we set $d=2$ and insert the anomalous phonon spectrum (\ref{spec-lowp}) in 
Eq. (\ref{rhon}) to obtain
\begin{equation}
\label{normal-dens}
\rho_{n0}^{\rm ph}
(T)\approx\frac{T^{[4-\sigma(T)]/\sigma(T)}}{8\pi^2}\frac{4-\sigma(T)}{\sigma(T)
\tilde{c}^{~4/\sigma(T)}}
\Gamma\left[\frac{4-\sigma(T)}{\sigma(T)}\right]
\zeta\left[\frac{4-\sigma(T)}{\sigma(T)}\right],
\end{equation}
where
$
\tilde{c}\equiv({g}_2n/m)^{1/2}a^{-\eta_0(T)}.$
By expanding  this
in powers of $T$, we obtain to leading order
the usual phonon contribution, which in
two dimensions is proportional to $T^3$. In deriving Eq. (\ref{normal-dens}) 
we have assumed the usual hydrodynamic limit where the upper cutoff --- here 
$p_0(T)$ --- is taken to be infinity. 

\section{Phase fluctuations}

By
integrating out small density fluctuations in the hydrodynamic 
limit, we obtain the
following effective action for the phase fluctuations:
\begin{equation}
\label{L-phase}
S_{\rm eff}= \int_0^\beta d\tau\left[
\frac{1}{2{g}_2}\int d^2r(\partial_\tau\theta)^2+H_{\rm eff}\right],
\end{equation}
where the effective Hamiltonian contains a local and a nonlocal
interaction between
the superfluid velocities
 ${\bf v}_s(\tau,{\bf r})=\nabla\theta(\tau,{\bf r})/m$:
\begin{equation}
H_{\rm eff}=H_{\rm eff}^{\rm local}+H_{\rm eff}^{\rm non-local},
\end{equation}
where
\begin{equation}
\label{Heff-local}
H_{\rm eff}^{\rm local}=\frac{mn}{2}\int d^2r{\bf v}_s^2(\tau,{\bf r}),
\end{equation}
and
\begin{equation}
\label{Heff}
H_{\rm eff}^{\rm non-local}= \frac{1}{2} \int d^2r\int d^2r'
{\cal M}({\bf r}-{\bf r}')~{\bf v}_s(\tau,{\bf r})\cdot{\bf v}_s(\tau,{\bf r}') ,
\end{equation}
with
\begin{equation}
\label{calG}
{\cal M}({\bf r}-{\bf r}')=\frac{mn(2a)^{-2\eta_0(T)}\Gamma[1-\eta_0(T)]}
{\pi\Gamma[\eta_0(T)]|{\bf r}-{\bf r}'|^{2[1-\eta_0(T)]}},
\end{equation}
being a bilocal mass density, which is obtained from the Fourier transform of 
${\cal M}(p)=nm(pa)^{-2\eta_0(T)}$. In the folowing discussion we will show that 
at high temperatures the non-local part contributes only in a small temperature 
interval above $T_c$. Below $T_c$ the usual effective Hamiltonian for 
the phase fluctuations given by Eq. (\ref{Heff-local}) 
dominates the critical behavior and the KT transition 
obtains. The arguments to be described below consider the scaling behavior 
of the spin wave theory for $H_{\rm eff}^{\rm non-local}$ and the field 
theory for the vortices, which consists of a generalized sine-Gordon theory. 

Let us consider first the case 
without the RPA correction, i.e., in the absence of the 
non-local effective Hamiltonian. This just corresponds to the usual phonon 
spectrum. In such a situation 
the effective Hamiltonian is given simply by Eq. (\ref{Heff-local}). The spin wave 
analysis is in this case well known. \cite{KT,Jose} However, it is useful to review it here in order 
to compare with the non-local spin wave regime.  
%
%

At higher temperatures we can neglect the higher Matsubara modes so that 
$\partial_\tau\theta=0$ and the effective action becomes 
\begin{equation}
S_{\rm eff}=\frac{n}{2mT}\int d^2r[\nabla\theta({\bf r})]^2,
\end{equation}
and the problem is essentially a classical one. Let us recall the computation of the 
correlation function $\langle\psi({\bf r})\psi^*({\bf r}')\rangle$ in this regime and 
in the absence of vortices (spin-wave theory). \cite{KT,Jose} 
In such a case we just have to compute the correlation between the phases:
\begin{equation}
\langle e^{i[\theta({\bf r})-\theta({\bf r}')]}\rangle
=\frac{1}{Z}\int{\cal D}\theta 
e^{-\int d^2 r''\left\{\frac{n}{2mT}[\nabla\theta({\bf r}'')]^2
+iJ({\bf r}'')\theta({\bf r}'')\right\}},
\end{equation}
where $J({\bf r}'')=\delta^2({\bf r}''-{\bf r})-\delta^2({\bf r}''-{\bf r}')$. The 
Gaussian integral is straightforward and yields
\begin{equation}
\label{correl-phs}
\langle e^{i[\theta({\bf r})-\theta({\bf r}')]}\rangle
=\exp\left\{\frac{mT}{n}[G({\bf r}-{\bf r}')-G(0)]\right\},
\end{equation}
where
\begin{equation}
G({\bf r})=\int\frac{d^2p}{(2\pi)^2}\frac{e^{i{\bf p}\cdot{\bf r}}}{p^2}.
\end{equation}
In order to evaluate the above Green function we introduce the regularized Green function
\begin{equation}
G_{M^2}({\bf r})=\int\frac{d^2p}{(2\pi)^2}\frac{e^{i{\bf p}\cdot{\bf r}}}{p^2+M^2},
\end{equation}
where $M$ is a regularizing mass to be sent to zero at the end of the calculations. 
Evaluating the integral explicitly, we obtain 
\begin{equation}
G_{M^2}({\bf r})=\frac{1}{2\pi}K_0(Mr),
\end{equation}
where $K_0(y)$ is the modified Bessel function of the second kind. On the other hand, we have 
\begin{equation}
G_{M^2}(0)=\frac{1}{2\pi}\ln\left(\frac{\Lambda}{M}\right),
\end{equation}
where $\Lambda=\pi/a$ is the ultraviolet cutoff. Now we can safely take the limit 
$M\to 0$:  
\begin{equation}
\label{2d-Green}
G({\bf r})-G(0)=\lim_{M\to 0}[G_{M^2}({\bf r})-G_{M^2}(0)]=-\frac{1}{2\pi}\ln\left(\frac{r}{a}\right)+{\rm const}.
\end{equation}
Therefore,
\begin{equation}
\langle e^{i[\theta({\bf r})-\theta({\bf r}')]}\rangle
\sim\frac{1}{|{\bf r}-{\bf r}'|^{\eta_0(T)}}.
\end{equation}
Note that precisely the exponent $\eta_0(T)$ that we have defined in Eq. (\ref{expon}) arises 
in the above equation. However, the mechanism that generates the anomalous behavior in the 
above classical spin-wave theory is completely different from the quantum case discussed 
in the previous Section. Indeed, there the anomalous scaling of the spectrum arises 
due to interaction effects, while in the above calculation it follows from the 
analytic properties of the Green function of a Gaussian classical theory in two dimensions. 

Let us now study the classical problem associated to the effective 
Hamiltonian at large distances. In this case 
(\ref{Heff}) dominates, since the corresponding power of $p$ is smaller than two. 
Once more we neglect the higher Matsubara modes to obtain the 
effective action for the classical problem as
\begin{equation}
S_{\rm eff}=\frac{1}{2T}\int d^2r\int d^2r'
{\cal M}({\bf r}-{\bf r}')~{\bf v}_s({\bf r})\cdot{\bf v}_s({\bf r}'), 
\end{equation}
where ${\bf v}_s({\bf r})\equiv{\bf v}_s(0,{\bf r})$. The correlation function 
between the phases has again the form (\ref{correl-phs}), except that the 
Green function is now given by
\begin{equation}
G({\bf r})=\int\frac{d^2p}{(2\pi)^2}\frac{e^{i{\bf p}\cdot{\bf r}}}{p^2{{\cal M}(p)}}.
\end{equation}
We have 
\begin{eqnarray}
G({\bf r})&=&\frac{a^{2\eta_0(T)}}{(2\pi)^2}\int_0^{2\pi}d\phi\int_0^\infty dp
\frac{e^{ipr\cos\phi}}{p^{1-2\eta_0(T)}}\nonumber\\
&=&\frac{a^{2\eta_0(T)}}{2\pi}\int_0^\infty dp\frac{J_0(pr)}{p^{1-2\eta_0(T)}},
\end{eqnarray}
where $J_0(y)$ is a Bessel function of the first kind. 
As before, we use a ultraviolet cutoff $\Lambda=\pi/a$ to evaluate $G(0)$:
\begin{equation}
G(0)=\frac{\pi^{2\eta_0(T)}}{4\pi\eta_0(T)}.
\end{equation}
We obtain finally
\begin{equation}
\label{2d-Green-anom}
G({\bf r})-G(0)=\frac{1}{4\pi}\left\{\left(\frac{2a}{r}\right)^{2\eta_0(T)}
\frac{\Gamma[\eta_0(T)]}{\Gamma[1-\eta_0(T)]}-\frac{\pi^{2\eta_0(T)}}{\eta_0(T)}\right\},
\end{equation}
which for small $\eta_0(T)$ can be rewritten as
\begin{equation}
G({\bf r})-G(0)\approx\frac{1}{4\pi\eta_0(T)}\left[\left(\frac{2a}{r}\right)^{2\eta_0(T)}
-1\right],
\end{equation}
and in the $\eta_0(T)\to 0$ limit Eq. (\ref{2d-Green}) is obviously reproduced. 

For arbitrary values of $0<\eta_0(T)<1$ it is not obvious to see how the KT theory 
is recovered when the anomalous phonon spectrum is taken into account. Recall that 
the logarithmic behavior in Eq. (\ref{2d-Green}) is crucial in the KT argument in the 
presence of vortices. Indeed, a simple scaling argument with free energy of the vortices 
combined with the results of spin-wave theory allows us to determine the value of 
$\eta_0(T)$ at the critical temperature $T_c$.\cite{KT} A more elaborate argument 
\cite{NelKost,Jose} using the RG shows that the stiffeness is 
renormalized and $\eta_0(T)$ becomes $\eta(T)=Tm^2/2\pi\rho_s(T)$. However, the value of 
$\eta(T)$ at $T_c$ is the same as the value of $\eta_0(T)$ at $T_c$. Such an RG analysis 
led to the cellebrated prediction of a universal jump of $\rho_s(T)$ as $T_c$ is 
approached from below.\cite{NelKost} At first sight we may think that an anomalous phonon 
spectrum would disrupt the whole argumentation due to the form of the 
corresponding non-local hydrodynamics (\ref{Heff}). In the following we will show that this 
{\it is not} the case. To this end it is necessary to proceed in two steps. First, 
we make an analysis about the validity of the hydrodynamic description given by (\ref{Heff}), which 
will provide us with a lower bound for $\eta_0(T)$. Second, we consider a careful analysis of 
the vortex field theory associated to this problem. We will see that the anomalous contribution 
to the vortex field theory becomes irrelevant at large distances and that, effectively, the 
{\it same} vortex field theory as in the KT case holds. This result will in turn provide us with 
a physical interpretation of the lower bound for $\eta_0(T)$.

Let us then start by discussing in more detail the effective Hamiltonian (\ref{Heff}). In order 
to define the mass density (\ref{calG}) we have to be able to Fourier transform 
${\cal M}(p)=nm(pa)^{-2\eta_0(T)}$. 
Such a Fourier transformation is 
only possible if $1/4<\eta_0(T)<1$. To see this we consider the integral
\begin{eqnarray}
\label{Fourier}
I(r)&=&\frac{1}{(2\pi)^2}\int_0^{2\pi}d\phi\int_0^\infty dp~\frac{p ~e^{ipr\cos\phi}}{
p^{2\eta_0(T)}}\nonumber\\
&=&\frac{1}{2\pi}\int_0^\infty dp\frac{J_0(pr)}{p^{2\eta_0(T)-1}}
\nonumber\\
&=&\frac{1}{2\pi ~r^{2[1-\eta_0(T)]}}\int_0^\infty dy\frac{J_0(y)}{y^{2\eta_0(T)-1}}.
\end{eqnarray}
The power of 
$r$ in the second line of Eq. (\ref{Fourier}) must be positive in order 
to make the above result well defined at large distances. This gives us 
once more the upper bound for $\eta_0(T)$, i.e., $\eta_0(T)<1$. The lower 
bound for $\eta_0(T)$ follows from the asymptotic behavior of the Bessel function 
for $y$ large,  
\begin{equation}
J_0(y)\sim\sqrt{\frac{2}{\pi y}}\cos(y-\pi/4).
\end{equation}
Thus, the integrand in Eq. (\ref{Fourier}) 
behaves for large $y$ like $\sim 1/y^{2\eta_0(T)-1/2}$ and it follows that 
in order to avoid a power-like ultraviolet divergence we must have 
$2\eta_0(T)-1/2>0$, which leads to   
$\eta_0(T)>1/4$. Therefore, in order to have a well defined mass density   
${\cal M}({\bf r}-{\bf r}')$ the inequality $1/4<\eta_0(T)<1$ has 
to be fulfilled. Note that 
the upper bound is associated to a large distance (infrared) divergence while the 
lower bound is to short distance (ultraviolet) divergence.
We have already seen that saturation of 
the upper bound leads to the determination of the temperature $T_*$. On the 
other hand, we will give arguments below showing  
that saturation of the lower bound determines the actual 
critical temperature of the system.  

At higher temperatures, near the phase transition, we are allowed to keep only
the zero Matsubara mode, such that the field theory becomes
two-dimensional. The vortices are introduced in the
standard way,\cite{Kleinert,NelKost} and a duality transformation
gives the following generalized sine-Gordon action
for the dual
field theory of vortices:
\begin{equation}
\label{SG}
S_{\rm dual}=\int d^2r\int d^2r'\left[\frac{T}{8\pi^2}
\Gamma^{-1}({\bf r}-{\bf r}')\partial_{\bf r}\varphi({\bf r})
\cdot\partial_{{\bf r}'}\varphi({\bf r}')
-z~\delta^2({\bf r}')\cos\varphi({\bf r})\right],
\end{equation}
where $\Gamma({\bf r}-{\bf r}')=mn\delta^2({\bf r}-{\bf r}')+
{\cal M}({\bf r}-{\bf r}')$ and 
$z$ is the fugacity of the gas of point vortices. A derivation 
of the action (\ref{SG}) is given in Appendix B. 
Nonlocal gradient terms in
sine-Gordon theory 
have  recently been discussed in
a different context,\cite{KNS,Herbut} where
a renormalization group (RG) analysis shows that
a local gradient term
$(\nabla\varphi)^2$
with a {\it positive} coefficient is generated.\cite{Herbut}
The nonlocal contributions can be
rewritten as a sum of higher-gradient terms which are all
irrelevant at large distances. Let us state this in more simple terms. In 
momentum space  
the non-local gradient has the form 
\begin{equation}
\frac{a^{2\eta_0(T)}T}{8\pi^2nm}\int\frac{d^2p}{(2\pi)^2}~
p^{2[1+\eta_0(T)]}\varphi({\bf p})\varphi(-{\bf p}).
\end{equation}
Thus, as a local gradient term is generated, 
we have that for small $p$ (large distances) the $p^2$-term dominates 
over the $p^{2[1+\eta_0(T)]}$ one 
and the anomalous contribution can be neglected. 
This argument shows that the dual theory  
(\ref{SG}) actually has a KT transition, since the fluctuation generated $p^2$-term 
is dominant in the infrared, leading effectively to a sine-Gordon theory of the usual type. 
Thus, our theory will ultimately be in agreement with the analysis of Ref. \onlinecite{Stoof} 
where no anomalous dimension arising from RPA corrections is considered.

In the low-temperature phase, where $0<\eta_0(T)\leq 1/4$, the hydrodynamic description via 
the effective Hamiltonian (\ref{Heff}) breaks down, since the inverse Fourier transform of 
${\cal M}(p)$ is no longer defined. This regime is equivalent to the one in the dual theory 
where the non-local gradient term becomes irrelevant. This means that the effective Hamiltonian 
of the classical theory in this range of temperatures is actually given by (\ref{Heff-local}). 
In other words, when the anomalous sine-Gordon theory  
becomes at large distances effectively the usual sine-Gordon theory 
(remember that a local gradient term is generated by fluctuations), we can dualize it back 
to obtain the {\it effective} original theory as given by Eq. (\ref{Heff-local}).  

In a KT transition 
the critical temperature $T_c$ is
determined from the equation  $\eta(T_c)=1/4$, or equivalently, 
$T_c=\pi\rho_s(T_c)/(2m^2)$.\cite{NelKost} 
Note that $\eta_0(T)$ corresponds 
to a low-temperature approximation to $\eta(T)$. Thus, we can  
interpret that saturating the lower bound in the spin-wave inequality 
$1/4<\eta_0(T)<1$ leads to an approximate value of the actual 
critical temperature for the KT transition. Therefore, we 
have that 
\begin{equation}
T_c\approx\frac{\pi\rho_s(0)}{2m^2}.
\end{equation}

We want to emphasize that 
the agreement of the value of $\eta_0(T)$ at the lower bound of the 
inequality $1/4<\eta_0(T)<1$ with the value 
$\eta(T_c)=1/4$ obtained from the KT theory is not a simple 
coincidence. It follows from 
the fact that the temperature at which $\eta_0(T)=1/4$ corresponds 
to the onset of the regime where the non-local term of the 
generalized sine-Gordon theory becomes irrelevant. 
 
We have obtained $\eta_0(T)$ as a correction 
to the power of the excitation spectrum at finite temperature by 
using a {\it classical} approximation to evaluate the density  
correlation function. 
Thus, we have simply accounted for a classical effect in a 
quantum calculation, i.e., our calculation of the spectrum at 
finite temperature is actually 
a semi-classical one. In the KT theory the value 
$\eta(T)=1/4$ is follows due to a conspiracy between spin-wave 
theory and the statistical mechanics of 
vortices. \cite{KT} In our case, a similar result obtains: our 
bound for $\eta_0(T)$ is derived by analysing the non-local spin-wave theory, 
while the conclusion that $\eta_0(T)=1/4$ determines the critical temperature 
needs additional analysis involving the vortex field theory (\ref{SG}). 
A more technical explanation follows by recalling that density and 
{\it longitudinal} phase 
fluctuations are related through the Ward identities. \cite{HM} 
More precisely, in Ref. \onlinecite{HM} it is shown that the 
density correlation function is related through 
an exact identity to the longitudinal component of the 
current correlation function, which corresponds to the response of the 
system to the longitudinal phase fluctuations or, in other words, spin-waves. 
Since $\eta_0(T)$ is 
determined by the density correlation function,
by taking the classical limit of this identity we obtain a relation between    
$\eta_0(T)$ and the spin-wave response. The vortices, on the other hand, 
are related to the {\it transverse} phase fluctuations. 
It can be shown\cite{HM} that the 
transverse component of the current correlation function decouples from 
all other correlation functions. This result holds also in the classical 
limit and 
that is the reason why the classical statistical 
mechanics of vortices alone can determine the phase transition.   

From the lower and upper bound to $\eta_0(T)$ we
obtain that the actual $T_c$ obtained from the 
KT theory is smaller than $T_*$. Indeed, we 
have that
\begin{equation}
T_c=\frac{\pi\rho_s(T_c)}{2m^2}<\frac{\pi\rho_s(0)}{2m^2}<
\frac{2\pi\rho_s(0)}{m^2}=T_*.
\end{equation}
Since $T_c<T_*$, 
it seems to exist a region $T_c<T<T_*$ where the system is not a 
superfluid, while not being a normal fluid either, due to the 
anomalous exponent. This region would  
be actually very small: from the inequality (\ref{ineq}) we see 
that a more correct statement would be to claim the existence 
of a temperature region $T_c<T\ll T_*$ where an anomalous normal 
state occurs. This state can be thought as a
quasi-condensate {\it without} phase coherence.
In this respect, it bears some resemblance with the pseudogap state in
high-$T_c$ superconductors,\cite{exp-sc} in which 
the phase fluctuations above the superconducting
critical temperature play a similar role.\cite{th-sc,KLB} 
In the pseudogap phase the spectrum is not like the one of a normal metal, 
although the system is not in a superconducting phase.  
However, we should warn the reader that the actual physics of the 
pseudogap state is likely much more complex and that such an analogy must 
be considered with utmost caution.

\section{Conclusion}

In this paper we have shown that temperature effects induce roton-like 
excitations in a dilute two-dimensional Bose gas. From this result it 
followed that for nonzero low temperatures the spectrum {\it is not} of the 
phonon type and has an anomalous scaling with temperature dependent 
exponent. Thus, we have obtained in the quantum regime a situation which is 
reminiscent of the Kosterlitz-Thouless transition, namely, a continuously 
varying exponent.   

It would be interesting to extend this analysis
to
strongly-coupled two-dimensional Bose systems such as films of $^4$He,
where experimental data have
 so far  been  fitted  by a
superfluid density whose vortex-independent background
contribution is calculated from the Bogoliubov spectrum.\cite{Reppy}
It was found
that the phonon contribution alone with
its $T^3$ behavior (recall that for a phonon spectrum 
we find in general a contribution behaving like 
$T^{d+1}$ in $d$ dimensions) is not consistent with the
temperature dependence of the data, thus calling for an improvement
of the theory.
In order to apply our approach to $^4$He films we
must
derive the $t$-matrix for
the actual atomic interaction in helium beyond the dilute limit.
Preliminary
work has been done some time ago,\cite{He-2d} and
phonon as well as roton excitations of the spectrum were obtained. However, these
works concentrate only on the low-temperature properties. We believe that 
thermally induced roton-like excitations should also occur in this case, though in a 
more complicate manner. 
Certainly, such strong-coupling
problems require more powerful
 calculation methods,
for example,
field-theoretic variational perturbation theory as developed 
in Ref. \onlinecite{SC}, which has led to the most accurate
predictions of critical exponents so far.\cite{LIPA} 

\acknowledgments

The authors would like to thank Axel Pelster 
and Anna Posazhennikova for many useful discussions.
This work was supported in part by the DFG Priority
Program SPP 1116, ``Interactions in Ultracold Atomic and Molecular Gases''.

\appendix \section{Classical limit of the polarization bubble}

Eq. (\ref{Pi}) can be rewritten 
as
\begin{equation}
\label{bubble-rw}
\widetilde \Pi(i\omega,{\bf p})=4m^2\int\frac{d^dq}{(2\pi)^d}n_B\left(
\frac{q^2}{2m}\right)\left(\frac{1}{2mi\omega-p^2-2{\bf p}\cdot
{\bf q}}-\frac{1}{2mi\omega+p^2-2{\bf p}\cdot{\bf q}}\right).
\end{equation}
In the classical approximation we write $n_b(x)\approx T/x$ and 
the polarization bubble can be rewritten as 
\begin{equation}
\label{picl}
\widetilde \Pi(i\omega,{\bf p})=4m^2T(I_+-I_-),
\end{equation}
where
\begin{equation}
I_\pm=-i\int\frac{d^dq}{(2\pi)^d}\frac{1}{2m\omega+i(2{\bf p}\cdot{\bf q}
\pm p^2)}\frac{1}{q^2}.
\end{equation}
The integrals $I_\pm$ can be evaluated using the Feynman parameters,  
\cite{KSF} 
\begin{equation}
I_\pm=-i\int_0^\infty d\lambda_1\int_0^\infty d\lambda_2 
\frac{d^dq}{(2\pi)^d}e^{-\lambda_1(2m\omega\pm ip^2+2i{\bf p}\cdot{\bf q})}
e^{-\lambda_2 q^2}.
\end{equation}
After evaluating the Gaussian integral over ${\bf q}$ we 
obtain
\begin{eqnarray}
I_\pm&=&-\frac{i}{(2\pi)^{d}}\int_0^\infty d\lambda_1\int_0^\infty d\lambda_2 
\left(\frac{\pi}{\lambda_2}\right)^{d/2}e^{-\lambda_1(2m\omega
\pm ip^2)} e^{-p^2\lambda_1^2/\lambda_2}\nonumber\\
&=&\mp\frac{(\pm i)^{d-2}}{2^d\pi^{d/2}}\Gamma(d/2-1)\Gamma(3-d)
p^{d-4}\left(1\mp\frac{2mi\omega}{p^2}\right)^{d-3}.
\end{eqnarray}
Substituting the above expression back into (\ref{picl}) we obtain 
Eq. (\ref{Pi1}). 

\section{Derivation of the anomalous sine-Gordon action}

At sufficiently high temperatures the Matsubara time dependence of the 
phase $\theta(\tau,{\bf r})$ can be neglected and the effective action 
(\ref{L-phase}) can be written simply as
\begin{equation}
S_{\rm eff}\approx\frac{1}{2T}\int d^2r\int d^2r'
{\cal M}({\bf r}-{\bf r}')~{\bf v}_s({\bf r})\cdot{\bf v}_s({\bf r}').
\end{equation}
Although the duality transformation can also be performed in the continuum, a 
more technically correct analysis is obtained in the lattice 
formalism. \cite{Kleinert} The lattice version of the above action suitable  
for a duality transformation is given by the Villain form
\begin{equation}
S_L=\frac{1}{2m^2T}\sum_{i,j}{\cal M}_{ij}(\nabla\theta_i-2\pi {\bf n}_i)
\cdot(\nabla\theta_j-2\pi {\bf n}_j),
\end{equation}
where we have set the lattice spacing to unit and the components 
of $\nabla\theta_i$ are understood as lattice derivatives. The 
field ${\bf n}_i$ is an integer field defined on the lattice. The 
partition function is then given by
\begin{equation}
Z=\sum_{\{{\bf n}_i\}}\int_{-\pi}^\pi\prod_i\frac{d\theta_i}{2\pi}
\exp(-S_L).
\end{equation}
The first step in the duality transformation is the introduction of 
an auxiliary field through a Gaussian completion, i.e.,
\begin{eqnarray}
&&\exp\left[-\frac{1}{2m^2T}\sum_{i,j}{\cal M}_{ij}(\nabla\theta_i-2\pi {\bf n}_i)
\cdot(\nabla\theta_j-2\pi {\bf n}_j)\right]
\nonumber\\
&\propto&\int_{-\infty}^\infty\prod_{k,\mu} db_{k\mu}
\exp\left[\frac{m^2T}{2}\sum_{i,j}{\bf b}_i({\cal M}^{-1})_{ij}
{\bf b}_j-{\rm i}\sum_j{\bf b}_j\cdot(\nabla\theta_j-2\pi {\bf n}_j)\right].
\end{eqnarray}
Next we apply the Poisson formula 
\begin{equation}
\sum_{n=-\infty}^\infty\int_{-\infty}^\infty dx f(x) e^{{\rm i}2\pi nx}
=\sum_{m=-\infty}^\infty f(m),
\end{equation}
in order to convert the integral over 
the real auxiliary ${\bf b}_i$ field into a sum over an integer field 
${\bf N}_i$. After this manipulation the periodic field $\theta_i$ can 
be easily integrated out after a summation by parts. This leads 
to the constraint $\nabla\cdot{\bf N}_i=0$. 
Up to unimportant overall factors the partition function becomes
\begin{equation}
Z=\sum_{\{{\bf N}_i\}}\delta_{\nabla\cdot{\bf N}_i,0}
\exp\left[-\frac{m^2T}{2}\sum_{i,j}{\bf N}_i({\cal M}^{-1})_{ij}
{\bf N}_j\right].
\end{equation}
In two dimensions the constraint is solved through
\begin{equation}
N_{i\mu}=\varepsilon_{\mu\nu}\nabla_\nu l_i.
\end{equation}
The resulting partition function will be the one of a neutral Coulomb gas in the 
lattice. By applying the Poisson formula once more, we obtain
\begin{equation}
Z= \sum_{\{s_i\}}\int_{-\infty}^\infty \prod_k d\varphi_k
\exp\left[-\frac{m^2T}{2}\sum_{i,j}\nabla_\mu\varphi_i({\cal M}^{-1})_{ij}
\nabla_\mu\varphi_j-{\rm i}\sum_j2\pi s_j \varphi_j\right].
\end{equation}
The integer fields $s_j$ represent the point vortices of the theory. 
We can now introduce the vortex 
fugacity $z$ to build a grand-canonical ensemble of vortices.\cite{Jose} The 
most relevant vortex configurations correspond to $s_j=\pm 1$. After 
the rescaling $\varphi_j\to\varphi_j/2\pi$ and taking the continuum 
limit of the grand-canonical ensemble theory, we obtain Eq. (\ref{SG}).

\end{document}